\documentclass{article}
\pdfoutput=1
\usepackage{graphicx}
\usepackage{amsmath}
\usepackage{amssymb}

\def\be{\begin{eqnarray}}
\def\ee{\end{eqnarray}}
\def\nn{\nonumber}

\def\tr{{\rm tr}\,}
\def\Tr{{\rm Tr}\,}

\def\l[{\phantom.[}

\hoffset= - 1.0in         

\begin{document}

\title{{\Large {\bf Universal Racah matrices and adjoint knot polynomials\\ Arborescent knots
}\vspace{.2cm}}
\author{{\bf A. Mironov$^{a,b,c,d,}$}\footnote{mironov@lpi.ru; mironov@itep.ru}, \ and \ {\bf A. Morozov$^{b,c,d,}$}\thanks{morozov@itep.ru},}
\date{ }
}

\maketitle

\vspace{-6.0cm}

\begin{center}
\hfill IITP/TH-17/15
\end{center}

\vspace{4.2cm}

\begin{center}
$^a$ {\small {\it Lebedev Physics Institute, Moscow 119991, Russia}}\\
$^b$ {\small {\it ITEP, Moscow 117218, Russia}}\\
$^c$ {\small {\it Institute for Information Transmission Problems, Moscow 127994, Russia}}\\
$^d$ {\small {\it National Research Nuclear University MEPhI, Moscow 115409, Russia }}
\end{center}

\vspace{1cm}

\begin{abstract}
By now it is well established that the
quantum dimensions of descendants of the adjoint representation
can be described in a universal form, independent of a particular
family of simple Lie algebras. The
Rosso-Jones formula then implies a universal description of the
adjoint knot polynomials for torus knots,
which in particular unifies the HOMFLY ($SU_N$) and Kauffman ($SO_N$) polynomials.
For $E_8$ the adjoint representation is also fundamental.
We suggest to extend the universality from the dimensions to the Racah matrices
and this immediately produces a unified description of
the adjoint knot polynomials for all arborescent (double-fat) knots,
including twist, 2-bridge and pretzel.
Technically we develop together the universality and
the "eigenvalue conjecture", which expresses the Racah and mixing matrices through the
eigenvalues of the quantum ${\cal R}$-matrix, and
for dealing with the adjoint polynomials one has to extend it to the previously unknown
$6\times 6$ case.
The adjoint polynomials do not distinguish between mutants and therefore
are not very efficient in knot theory, however, universal polynomials
in higher representations can probably be better in this respect.
\end{abstract}

\vspace{.5cm}

\bigskip

\section{Introduction}

Knot theory \cite{knotpols}-\cite{W1} is intimately connected to representation theory
via the Reshetikhin-Turaev (RT) formalism \cite{RT}-\cite{knotebook1}.
For the simplest (torus) knots the Rosso-Jones formula \cite{RJ}-\cite{DMMSS} expresses
knot invariants through just quantum dimensions and Casimir eigenvalues.
For evaluating knot polynomials of other knots, one has to know the Racah matrices.
In fact, for a broad family of arborescent (double-fat) knots \cite{Con,Cau,BS} it is sufficient to know only two Racah
matrices \cite{inds,gmmms,mmmrv}. Moreover, in the case of self-contragredient representation they are proportional to each other
so that one needs only one Racah matrix, and this is exactly the case of adjoint representation which we consider in this paper.
However, in general also needed are the "mixing matrices" \cite{MMMkn2,MM3f2},
which are contractions of several Racah so that to construct them one has to know a series of Racah matrices.

A remarkable discovery in representation theory is {\it universality}
\cite{Kostant}-\cite{RM},
a possibility to describe quantities for all the simple Lie algebras
by a universal formula, which is the same for unitary, orthogonal,
symplectic and exceptional groups.
It turns out that such universal description exists, provided one
considers only adjoint representation and its descendants
(representations, appearing in tensor powers of adjoint),
instead of arbitrary descendants of the fundamental one.

Historically, the term "universality" refers to the notion of the
"Universal Lie algebra", introduced by Vogel in \cite{V2},
which, roughly speaking, was intended to describe the $\Lambda$-algebra of triple-ended Feynman diagrams
(closely related to Connes-Kreimer description \cite{CK,GMS}). These diagrams are related to the Vassiliev invariants and, on the physical side, to the perturbative expansion in Chern-Simons theory, and the universality came just as an observation, as it was also the case earlier \cite{SOSp}.
We call some quantity in the theory of (quantum) simple Lie algebras universal,
if it can be expressed as a smooth symmetric function of
three parameters $u=q^\alpha,\ v=q^\beta,\ w=q^\gamma$,
and takes values for a given simple Lie algebra at the corresponding points of Vogel's Table,
see (\ref{VogelT}) below in this text
(for the ADE case it appeared already in \cite{Kostant}).

Recently in \cite{MMkrM} it was suggested to extend the notion of universality to
the adjoint knot polynomials,
i.e. to express them in terms of three variables $u,\ v,\ w$
instead of the more conventional two $A,\ q$. Conventional colored polynomials (in the "$E_8$-sector" of representation theory) appear on particular $2$-dimensional slices of the $(u,v,w)$ space. For instance, the $SU(N)$ (HOMFLY) and
$SO(N)$ (Kauffman) polynomials are described by choosing $u=q^{-2},\ v=q^2,\ A=q^N$ and $u=q^{-2},\ v=q^4,\ A=q^{N-1}$ respectively. However, in terms of
$u,\ v,\ w$ the knot polynomials acquire an additional property: they are symmetric functions of these variables.
In \cite{MMkrM} such universal formulas were explicitly presented for a variety of knots that
included 2-and 3-strand torus knots and links and the figure eight knot, for the 2-strand case
they were later reproduced in \cite{West}.
Actually, for the torus case this is not a big surprise, because the Rosso-Jones formula
providing the knot polynomials in this case does not contain anything but quantum dimensions
and Casimir eigenvalues, which are known to possess universal description \cite{V2,Vogel,LM,MV12,RMD}.
Still , the results of \cite{MMkrM} were originally obtained without any use
of the Rosso-Jones formula, moreover, in the figure eight case this formula is unapplicable at all.
The actual message of  \cite{MMkrM}, which we make explicit in the present and the sequel paper \cite{II},
is that {\bf the universality can be lifted to the Racah and mixing matrices}, and thus all the
knot polynomials in adjoint family can actually be represented in the universal form.
In this paper we make this idea explicit for the entire family of arborescent
knots (which can be presented by double-fat graphs), evaluation of their knot polynomials having been discussed in detail in \cite{mmmrv}.

We achieve this by developing another challenging conjecture:
the eigenvalue conjecture (EC) of \cite{IMMMev} (see also \cite{evmath})
expressing the mixing matrices between ${\cal R}$-matrices acting on different
pairs of adjacent strands in a braid through the eigenvalues
of ${\cal R}$-matrices themselves (actually, those eigenvalues are $\lambda=q^\kappa$ with $\kappa$ being value of the second Casimir
operator).

Actually, the EC in \cite{IMMMev} is formulated for the 3-strand closed braids, and then the mixing matrices are actually the Racah matrices. At the same time, once the Racah matrix is known, ideas of \cite{mmmrv} can be used to evaluate the colored polynomials for the arborescent knots, which is a family very different from the 3-strand one (sometimes wider, sometimes narrower). In fact, the 3-strand family {\it per se} can also be studied by the method of the present paper, but this is a separate story that will be reported elsewhere \cite{II}. Technically EC provides a solution to the Yang-Baxter equation
\be
{U}{\cal R}{U}{\cal R}{U} = {\rm diagonal}
\label{YB}
\ee
with diagonal ${\cal R} = {\rm diag}(\lambda_1,\ldots,\lambda_{M})$
in the form of orthogonal matrix $U$,
\be
U^{\tr} = U, \ \ \ \ \ U^2 = I
\label{Usquare}
\ee
with all entries $U_{ij}$ explicitly expressed through the eigenvalues $\lambda$'s
(the number of Yang-Baxter equations is exactly equal to the number of independent
angles in the orthogonal matrix).
In \cite{IMMMev} such solution was explicitly provided for $M\leq 5$,
for the purposes of the present paper we need an extension to $M=6$,
what is not at all straightforward (as emphasized also in \cite{evmath}).

Note that in terms of representation theory, $U$ is constructed as the Racah matrix that relates the two maps:
$\underbrace{(V\otimes V)}_Q\otimes V\to W$ and $V\otimes\underbrace{(V\otimes V)}_{Q'}\to W$ (for the sake of simplicity, we discuss
here only the case of knots, not links when one is not obliged to consider three coinciding representations $V$). This matrix
is involved in evaluating the knot invariants from 3-strand braid representation. At the same time, evaluating the arborescent knot
invariants involves the Racah matrices $\bar S$ and $S$ that relate accordingly the maps
$\underbrace{(V\otimes \bar V)}_Q\otimes V\to V$ with $V\otimes\underbrace{(\bar V\otimes V)}_{Q'}\to V$
and $\underbrace{(\bar V\otimes V)}_Q\otimes V\to V$ with $\bar V\otimes\underbrace{(V\otimes V)}_{Q'}\to V$, where $\bar V$ denotes the
representation contragredient to $V$. In the case of adjoint representation $V$, which is self-contragredient, all three matrices
$U$, $S$ and $\bar S$ are equal to each other and the property of one of them are immediately inherited by the remaining ones. In particular, since this case is multiplicity free, all three matrices are symmetric (due to symmetricity of $\bar S$), and their first row is (the property inherited from $S,\ \bar S$)
\be\label{dim}
U_{1Q}=U_{\emptyset Q}={\sqrt{{\cal D}_Q}\over {\cal D}_{Adj}}
\ee
where ${\cal D}_Q$ denotes the quantum dimension of the representation $Q$.

What we do in the present paper, we solve two problems at once:
we use the coincidence of these Racah matrices in order to apply the EC to $U$, additionally using its symmetricity and the form of the first row. This allows us to restore the matrix $U$ and then we use $S=U$ to construct the universal adjoint polynomials of the arborescent knots.

The paper is organized as follows. In section 2 we construct a $6\times 6$ mixing matrix that is further used in section 3 for constructing the universal adjoint knot polynomials of various arborescent knots. Section 4 contains some comments of extension from the family of arborescent knots and also a discussion of properties of the polynomials obtained in the present paper. The concluding remarks are in section 5.

This paper is a the first paper of series of two papers devoted to evaluating the universal adjoint polynomials. The second paper \cite{II} contains the results for more general knots that can be presented by "fingered 3-strand closed braids" \cite{MM3f1,MM3f2}.

\section{$6\times 6$ mixing matrix}

\subsection{Generality\label{mima}}

In accordance with what is said in the Introduction, we need to construct the Racah matrix $U_{QQ'}$ that maps
$\underbrace{(Adj\otimes Adj)}_Q\otimes Adj\to Adj$ and $Adj\otimes\underbrace{(Adj\otimes Adj)}_{Q'}\to Adj$.
Since
\be
Adj^{\otimes 2} = [\emptyset] \oplus Y_2  \oplus Y_2' \oplus Y_2'' \oplus Adj \oplus X_2
\label{decoAdj2}
\ee
is decomposed into six irreps (they are actually {\it ir}reducible with respect to a Lie
algebra, multiplied by a discrete symmetry of Dynkin diagram),
the relevant Racah matrix is $6\times 6$, as it was already claimed. Now we are going to apply the EC.

To understand the structure of eigenvalue formulas, it deserves looking at
the $2\times 2$ case.
Generic symmetric orthogonal matrix --solution to (\ref{Usquare}) in this case is just
\be
U=\left(\begin{array}{cc} c & s \\ s & - c\end{array}\right)
\ee
with $c^2+s^2=1$.
Substitution into (\ref{YB}) gives
\be
(\lambda_1^2-\lambda_1\lambda_2+\lambda_2^2)c^2 + \lambda_1\lambda_2s^2  = 0 \ \
\Longrightarrow \ \
U=\frac{1}{\lambda_1-\lambda_2} \left(\begin{array}{cc}
\sqrt{-\lambda_1\lambda_2} & \sqrt{\lambda_1^2-\lambda_1\lambda_2+\lambda_2^2} \\
\sqrt{\lambda_1^2-\lambda_1\lambda_2+\lambda_2^2} & - \sqrt{-\lambda_1\lambda_2}\end{array}\right)
\ee
This structure is inherited by the mixing matrices for higher $M>2$.
It involves a parameter $T^2 = \prod_{m=1}^M \lambda_m$ and different expressions
for diagonal and the squares of the off-diagonal elements of $U$, with simple denominators
and some $M$-dependent polynomials of $\lambda$ and $T_M$ in the numerators.

In particular, the mixing matrix for $M=6$ can be obtained using the coincidence of matrices $U$, $S$ and $\bar S$ as was explained in the Introduction, which adds to the properties of mixing matrix also the properties of $S$ and $\bar S$.
The result reads:

\noindent
diagonal elements:
\be\label{diag66}
U_{ii} = \frac{\lambda_i^2}{\prod_{m\neq i}^6 (\lambda_i-\lambda_m)}\left(
\lambda_i\sum_{\stackrel{m<n}{m,n\neq i}}\lambda_m\lambda_n -
\sum_{\stackrel{l<m<n}{l,m,n\neq i}}\lambda_l\lambda_m\lambda_n\ \
+\ \ T\cdot\!\sum_{m\neq i} \frac{\lambda_i^2-\lambda_m^2}{\lambda_i\lambda_m}\right)
\ee
squares of the off-diagonal elements:
\be
U_{ij}^2 = \frac{\sqrt{\lambda_i} (\lambda_i^3-T)}{T\cdot \prod_{m\neq i}^6(\lambda_i-\lambda_m)}\cdot
\frac{\sqrt{\lambda_j} (\lambda_j^3-T)}{T\cdot \prod_{m\neq j}^6(\lambda_j-\lambda_m)}\cdot
\prod_{\stackrel{m<n}{m,n\neq i,j}}
\Big(\sqrt{\lambda_i\lambda_m\lambda_n}+ \sqrt{\lambda_j\lambda_{m'}\lambda_{n'}}\Big)
\label{offdiag66}
\ee
and
$T^2 = \prod_{m=1}^6 \lambda_m = (\lambda_i\lambda_m\lambda_n)\cdot( \lambda_j\lambda_{m'}\lambda_{n'})$
where $m'$ and $n'$ are the complements of $\{i,j,m,n\}$ in the set $\{1,2,\ldots,6\}$.

To correctly deal with the matrix elements (\ref{diag66}), (\ref{offdiag66}), one has choose properly the signs of roots
depending on signs of the eigenvalues $\lambda_i$. Now we come to the concrete case of the adjoint representation.

\subsection{Tensor square of adjoint representation}

As we already wrote, the tensor square of adjoint representation is given by formula (\ref{decoAdj2}).
The last two representations $Adj$ and $X_2$ belong to the antisymmetric square
$\Lambda^2(Adj)$, while the first four to the symmetric square ${\cal S}^2(Adj)$.
The corresponding ${\cal R}$-matrix eigenvalues, which are equal to $\lambda_Q=q^{\kappa_Q}$ with $\kappa_Q$ being the second Casimir, are
\be\label{eva}
\lambda_\emptyset = 1, \ \ \ \lambda_{Y_2} = uv^2w^2, \ \ \
 \lambda_{Y_2'} = vu^2w^2, \ \ \  \lambda_{Y_2''} = wu^2v^2, \ \ \
  \lambda_{Adj} = -uvw, \ \ \  \lambda_{X_2} = -(uvw)^2
\ee
The quantum dimensions that makes a content of the first row of the mixing matrix, (\ref{dim}) are
\be
{\cal D}_{Adj} =   -\frac{\{\sqrt{u}vw\}\{\sqrt{v}uw\}\{\sqrt{w}uv\}}{\{\sqrt{u}\}\{\sqrt{v}\}\{\sqrt{w}\}}
\ee
\be
{\cal D}_{Y_2} =
\frac{\{uvw\}\{u\sqrt{v}w\}\{uv\sqrt{w}\}\{v\sqrt{uw}\}\{w\sqrt{uv}\}\{vw/\sqrt{u}\}}
{\{\sqrt{u}\}\{u \}\{\sqrt{v}\}\{\sqrt{w}\}\{\sqrt{u/v}\}\{\sqrt{u/w}\}}
\ee
\be
{\cal D}_{X_2} = -\ {\cal D}_{Adj}\cdot\frac{\{u\sqrt{vw}\}\{v\sqrt{uw}\}\{w\sqrt{uv}\}}{\{u\}\{v\}\{w\}}
\left(\sqrt{uv}+\frac{1}{\sqrt{uv}}\right)\left(\sqrt{vw}+\frac{1}{\sqrt{vw}}\right)
\left(\sqrt{uw}+\frac{1}{\sqrt{uw}}\right)
\label{dimssquare}
\ee
and the quantum dimensions for $Y_2'$ and $Y_2''$ are obtained from $Y_2$ by cyclic permutations of the triple $(u,v,w)$.
These representations for the concrete case of $SU(N)$ are
\be
Y_2 = [42^{N-2}], \ \ Y_2' =[2^21^{N-4}], \ \ Y_2''=Adj,\ \ X_2=[31^{N-3}]\oplus [3^22^{N-3}]
\ee
where we extended the $SU(N)$ group by automorphisms of its Dynkin diagram, so that the sum of two last representations becomes one irreducible representation of the extended group. For $SO(N)$ these representations are
\be
Y_2 = [22], \ \ Y_2' = [1111], \ \ Y_2''=[2],\ \ X_2=[211]
\ee

\bigskip

Dimensions, eigenvalues, Racah/mixing matrices and knot
polynomials for particular Lie algebras arise under substitutions of
$u=q^\alpha$, $v=q^\beta$, $w=q^\gamma$ from Vogel's table

\be
\begin{array}{|c|c|c|c|}
\hline   
\text{algebra} & \alpha & \beta & \gamma  \\
 \hline  
SU(N) &    -2 & 2 & N \\
 \hline  
SO(N) & -2 & 4 & N-4 \\
 \hline
Sp(N)  & -2 & 1 & \hbox{\small {$\frac{1}{2}$}}N+2\\
 \hline  
Exc(N) & -2 & N+4 & 2N+4 \\
 \hline
\end{array}
\label{VogelT}
\ee
where all exceptional simple Lie algebras belong to the $Exc$ line at special values of parameter:

\be
\begin{array}{|c||c|c|c|c|c|c|c|c|c|c|c|c|}
\hline 
N & -1  & -2/3 &0 & 1 & 2 & 4 & 8 \\
\hline  
Exc(N) & A_2 & G_2& D_4 & F_4 & E_6& E_7 & E_8 \\
\hline
\end{array}
\label{VogelExc}
\ee

\subsection{The universal Racah matrix for adjoint}

Now are finally ready to write down the $6\times 6$ mixing matrix for the concrete eigenvalues of the adjoint (\ref{eva}) which we further use for constructing the universal knot polynomials for the arborescent knots. In the latter case we use the standard notation
for the Racah matrix $S$, hence the notation $U={\cal S}$ below.

The matrix is symmetric and independent items are ($t=uvw$):
\be
{\cal S}_{11} = -\frac{(u-1)(v-1)(w-1)t^3}{(t^2-u)(t^2-v)(t^2-w)}
\nn\\
{\cal S}_{22} = \frac{u^2}{(t^2-u)(u-v)(u-w)(u+1)(t+u)}\cdot
\Big(-u-t+u(v+w)(1+t)-vw(1+2t)+t(v+w)(1+vw)-\nn \\
-u^2(v+w)t-t^2(v+w)(u+1)+ut^2(2+t)+t^3(v-1)(w-1)\Big)
\nn\\
{\cal S}_{33} = \frac{v^2}
{(t^2-v)(v-u)(v-w)(v+1)(t+v)}\cdot
\Big(-v-t+v(u+w)(1+t)-uw(1+2t)+t(u+w)(1+wu)-\nn\\
-v^2t(u+w)-t^2(u+w)(v+1)+vt^2(2+t)+t^3(u-1)(w-1)\Big)
\nn
\ee
\be
{\cal S}_{44} = \frac{w^2}{(t^2-w)(w-u)(w-v)(w+1)(t+w)}\cdot
\Big(-w-t+w(u+v)(1+t)-uv(1+2t)+t(u+v)(1+uv)-\nn\\
-w^2t(u+v)-t^2(u+v)(w+1)+wt^2(2+t)+t^3(u-1)(v-1)\Big)
\nn\\
{\cal S}_{55} = -\frac{(uv+wu+vw)(1-2t)+t(3-2(u+v+w))+t^2(u+v+w-3)}{(t-1)(1+uv)(1+wu)(1+vw)}
\nn\\
{\cal S}_{66} = -\frac{1+uv+wu+vw-t(u+v+w)-t^2}{t-1)(u+1)(v+1)(w+1)}
\nn\\
{\cal S}_{12} = \frac{1}{t^2-u}\cdot\sqrt{\frac{ut(v-1)(w-1)(t^2-u^3)(vt-1)(wt-1)(t^2-1)}
{(u+1)(u-v)(u-w)(t^2-v)(t^2-w)}}
\nn\\
{\cal S}_{13} = \frac{1}{t^2-v}\cdot\sqrt{\frac{vt(u-1)(w-1)(t^2-v^3)(ut-1)(wt-1)(t^2-1)}
{(v+1)(v-u)(v-w)(t^2-u)(t^2-w)}}
\nn\\
{\cal S}_{14} = \frac{1}{t^2-w}\cdot\sqrt{\frac{wt(u-1)(v-1)(t^2-w^3)(ut-1)(vt-1)(t^2-1)}
{(w+1)(w-u)(w-v)(t^2-u)(t^2-v)}}
\nn\\
{\cal S}_{15} = uvw\cdot\sqrt{-\frac{t(u-1)(v-1)(w-1)}{(t^2-u)(t^2-v)(t^2-w)}}
\nn\\
{\cal S}_{16} = \sqrt{\frac{t(1+uv)(1+wu)(1+vw)(ut-1)(vt-1)(wt-1)}
{(u+1)(v+1)(w+1)(t^2-u)(t^2-v)(t^2-w)}}
\nn\\
{\cal S}_{23} = \frac{1}{u-v}\cdot\sqrt{-\frac{(u-1)(v-1)(ut-1)(vt-1)(t^2-u^3)(t^2-v^3)}
{uv(u+1)(v+1)(u-w)(v-w)(t^2-u)(t^2-v)}}
\nn\\
{\cal S}_{24} = \frac{1}{u-w}\cdot\sqrt{-\frac{(u-1)(w-1)(ut-1)(wt-1)(t^2-u^3)(t^2-w^3)}
{uw(u+1)(w+1)(u-v)(w-v)(t^2-u)(t^2-w)}}
\nn\\
{\cal S}_{34} = \frac{1}{v-w}\cdot\sqrt{-\frac{(v-1)(w-1)(vt-1)(wt-1)(t^2-v^3)(t^2-w^3)}
{vw(v+1)(w+1)(w-u)(v-u)(t^2-v)(t^2-w)}}
\nn\\
\nn\\
{\cal S}_{25} = -\frac{1}{vw+1}\cdot\sqrt{-\frac{(u-1)(t^2-u^3)(vt-1)(wt-1)(t+1)}{u(u+1)(u-v)(u-w)(t^2-u)(t-1)}}
\nn\\
{\cal S}_{35} = \frac{1}{uw+1}\cdot\sqrt{-\frac{(v-1)(t^2-v^3)(ut-1)(wt-1)(t+1)}{v(v+1)(v-u)(v-w)(t^2-v)(t-1)}}
\nn\\
{\cal S}_{45} = -\frac{1}{uv+1}\cdot\sqrt{-\frac{(w-1)(t^2-w^3)(ut-1)(vt-1)(t+1)}{w(w+1)(w-u)(w-v)(t^2-w)(t-1)}}
\nn\\
\nn\\
{\cal S}_{26} = -\frac{1}{u+1}\cdot\sqrt{\frac{(v-1)(w-1)(uv+1)(uw+1)(t^2-u^3)(ut-1)(t+1)}
{u(v+1)(w+1)(u-v)(u-w)(vw+1)(t^2-u)(t-1)}}
\nn\\
{\cal S}_{36} = -\frac{1}{v+1}\cdot\sqrt{\frac{(u-1)(w-1)(uv+1)(vw+1)(t^2-v^3)(vt-1)(t+1)}
{v(u+1)(w+1)(v-u)(v-w)(uw+1)(t^2-v)(t-1)}}
\nn\\
{\cal S}_{46} = -\frac{1}{w+1}\cdot\sqrt{\frac{(u-1)(v-1)(uw+1)(vw+1)(t^2-w^3)(wt-1)(t+1)}
{w(u+1)(v+1)(w-u)(w-v)(uv+1)(t^2-w)(t-1)}}
\nn\\
\nn\\
{\cal S}_{56} = \frac{1}{t-1}\cdot
\sqrt{-\frac{(u-1)(v-1)(w-1)(ut-1)(vt-1)(wt-1)}{(u+1)(v+1)(w+1)(uv+1)(uw+1)(vw+1)}}
\label{uniRacah}
\ee
In practical calculations one needs also matrix elements like
\be
\Big({\cal S}{\cal R}^p{\cal S}\Big)_{ij} = \sum_{k=1}^6 {\cal S}_{ik}\lambda_k^p {\cal S}_{kj}
\ee
which are (partly) listed in \cite{knotebook2}.

\section{Universal adjoint knot polynomials}

\subsection{Mixing matrix and torus knots: checks of consistency}

By construction, the matrix (\ref{uniRacah}) possesses all the necessary properties: it is symmetric, orthogonal (thus, its square is equal to 1), satisfies the Yang-Baxter equation (\ref{YB}). Here we check that it gives correct contributions to the universal adjoint polynomials of the 2- and 3-strand torus knots and links earlier obtained in \cite{MMkrM}. In the 2-strand case these are given by
\be
{\cal P}_{Adj}^{[2,n]} = (uvw)^{4n}{\cal D}_{Adj}\cdot \sum_{Q=1}^6 {\cal S}_{1 Q}^2\lambda_Q^{-n}
\label{Padj2str}
\ee
and, using  formulas (\ref{decoAdj2}) and (\ref{eva}), we get
\be
{\cal P}_{Adj}^{[2,n=2k+1]} = \frac{(uvw)^{4n}}{{\cal D}_{Adj}}\Big(1
+\frac{u^n}{(uvw)^{2n}}{\cal D}_{Y_2}
+ \frac{v^n}{(uvw)^{2n}}{\cal D}_{Y_2'}
+\frac{w^n}{(uvw)^{2n}}{\cal D}_{Y_2''}
- \frac{1}{(uvw)^n}{\cal D}_{Adj} - \frac{1}{(uvw)^{2n}}{\cal D}_{X_2}\Big)
\ee
and
\be
{\cal P}_{Adj}^{[2,n=2k]} = \frac{(uvw)^{4n}}{{\cal D}_{Adj}}\Big(1
+\frac{u^n}{(uvw)^{2n}}{\cal D}_{Y_2}
+ \frac{v^n}{(uvw)^{2n}}{\cal D}_{Y_2'}
+\frac{w^n}{(uvw)^{2n}}{\cal D}_{Y_2''}
+\frac{1}{(uvw)^n}{\cal D}_{Adj} +\frac{1}{(uvw)^{2n}}{\cal D}_{X_2}\Big)
\ee
in full accord with \cite{MMkrM}.

In the 3-strand case we can either use the universal Rosso-Jones formula
from \cite{MMkrM} or derive it from the mixing matrices:
\be\label{3st}
{\cal P}^{[3,n]}_{Adj} = \frac{(uvw)^{8n}}{{\cal D}_{Adj}}\cdot
\sum_{Q\in Adj^{\otimes 3}} c_Q \lambda_Q^{-2n/3}{\cal D}_Q
=  \frac{(uvw)^{8n}}{{\cal D}_{Adj}}\cdot
\sum_{Q\in Adj^{\otimes 3}} \Tr_{W_Q} ({\cal R}{\cal S}{\cal R}{\cal S})_Q^{-n}
\ee
In other words, the coefficients $c_Q$, which depend on $n$ only modulo $3$,
can be extracted either from the Adams rule or evaluated from traces over the
intertwiner spaces $W_Q$ in $Adj^{\otimes 3} = \oplus_Q W_Q\otimes Q$ at the r.h.s.
To do this, we need to know the eigenvalues of $\Big({\cal R}{\cal S}{\cal R}{\cal S}\Big)_Q$,
where eigenvalues of ${\cal R}$ are quantum Casimirs of representations from $Adj^{\otimes 2}$,
leading to $Q \in Adj^{\otimes 3}$,
and mixing matrices ${\cal S}$ are made out of them by
the eigenvalue conjecture of \cite{IMMMev}.

In particular,

$Q=\emptyset$ comes from $X_1=Adj \in Adj^{\otimes 2}$, the corresponding $W_\emptyset$
is one dimensional and
\be\label{t0}
\Big({\cal R}{\cal S}{\cal R}{\cal S}\Big)_\emptyset = \lambda_{X_1}^2 = (uvw)^2
\ee

For $Q=X_1$ the space $W_{X_1}$ is 6-dimensional, the corresponding mixing matrix is exactly
(\ref{uniRacah}), and
\be\label{t1}
e.v\Big\{({\cal R}{\cal S}{\cal R}{\cal S})_{X_1}\Big\} =
(uvw)^{8/3}\cdot \{1,e^{2\pi i/3}, e^{-2\pi i/3},1,e^{2\pi i/3}, e^{-2\pi i/3}\}
\ee

For $Q=Y_2$ the space $W_{Y_2}$ is 3-dimensional, and
\be\label{t2}
e.v\Big\{({\cal R}{\cal S}{\cal R}{\cal S})_{Y_2}\Big\} = (uvw)^{10/3}u^{-2/3}\cdot
  \{1,e^{2\pi i/3}, e^{-2\pi i/3}\}
\ee
and so on for other representations $Q\in Adj^{\otimes 3}$.

In result neither $X_1=Adj$ nor $Y_2$ contributes to the universal adjoint polynomials of the 3-strand torus knots,
when $n=3k\pm 1$, but both contribute in the case of links, when $n=3k$,
in full accordance with the answers in \cite{MMkrM}:
\be\label{3sk}
{\cal P}^{[3,n=3k\pm 1]}_{\rm Adj} = \frac{(uvw)^{4n}}{{\cal D}_{Adj}}\cdot\Big((uvw)^{2n}+ {\cal D}_{X_3}
+ u^{2n} {\cal D}_{Y_3 } + v^{2n} {\cal D}_{Y_3' } + w^{2n} {\cal D}_{Y_3''}
-u^{n} {\cal D}_{C} - v^{n} {\cal D}_{C'} - w^{n} {\cal D}_{C''}
\Big)
\ee
and
\be\label{3sl}
{\cal P}^{[3,n=3k]}_{\rm Adj} = \frac{(uvw)^{12k}}{{\cal D}_{Adj}}\cdot
\Big((uvw)^{6k}+6(uvw)^{4k}{\cal D}_{X_1} +6(uvw)^{2k}{\cal D}_{X_2}
+{\cal D}_{{\cal X}_3}+
 u^{6k} {\cal D}_{Y_3 } + v^{6k} {\cal D}_{Y_3'} + w^{6k} {\cal D}_{Y_3''} + \nn \\
+3(vw)^{2k}{\cal D}_{B }+3(uw)^{2k}{\cal D}_{B'}+3(uv)^{2k}{\cal D}_{B''}
+2u^{3k} {\cal D}_{C }+2v^{3k} {\cal D}_{C'} + 2w^{3k} {\cal D}_{C''}+\nn \\
+3u^{2k}(uvw)^{2k}{\cal D}_{Y_2 }
+3v^{2k}(uvw)^{2k}{\cal D}_{Y_2'} +3w^{2k}(uvw)^{2k}{\cal D}_{Y_2''}
\Big)
\ee

Explicit expressions for dimensions and Casimirs, in addition to those
listed in (\ref{t0})-(\ref{t2}), are \cite{RMD}:
\be
{\cal D}_{Y_3 } = -\frac{\{uvw\}\{v\sqrt{w}\}\{w\sqrt{v}\} \{v\sqrt{uw}\}\{w\sqrt{uv}\}
\{uv\sqrt{w}\}\{uw\sqrt{v}\}\{vw/u\sqrt{u}\}\{vw\sqrt{u}\}}
{\{\sqrt{u}\}\{\sqrt{v}\}\{\sqrt{w}\} \{u\}\{u\sqrt{u}\}\{\sqrt{v}/u\}
\{\sqrt{w}/u\}\{\sqrt{u/v}\}\{\sqrt{u/w}\}}   \nn\\
{\cal D}_{B } = -\frac{\{uvw\}\{v\sqrt{uw}\}\{w\sqrt{uv}\}\{uv\sqrt{w}\}\{uw\sqrt{v}\}\{vw\sqrt{u}\}
\{u\sqrt{v}\}\{u\sqrt{w}\}\{uv/\sqrt{w}\}\{uw/\sqrt{v}\}}{\{\sqrt{u}\}\{u\}\{\sqrt{v}\}^2\{\sqrt{w}\}^2
\{\sqrt{v}/w\}\{\sqrt{w}/v\}\{\sqrt{v/u}\}\{\sqrt{w/u}\}} \nn \\
{\cal D}_{C } =
-\frac{\{uvw\}\{vw\}\{v\sqrt{w}\}\{w\sqrt{v}\}\{u\sqrt{vw}\}\{uv\sqrt{w}\}\{uw\sqrt{v}\}\{v\sqrt{uw}\}\{w\sqrt{uv}\}}
{\{\sqrt{u}\}^2\{u^{3/2}\}\{\sqrt{v}\}\{\sqrt{w}\}\{\sqrt{u/v}\}\{\sqrt{u/w}\}\{\sqrt{u}/v\}\{\sqrt{u}/w\}}\cdot
\nn  \\
\cdot\left(\sqrt{uv}+\frac{1}{\sqrt{uv}}\right)\left(\sqrt{uw}+\frac{1}{\sqrt{uw}}\right)
\left(\sqrt{\frac{u}{vw}}+\sqrt{\frac{vw}{u}}\right)
\ee
and
\be
\lambda_{X_i} = (uvw)^i,  \ i=0,1,2,3, \ \ \ \
\lambda_{Y_2}  = uv^2w^2, \ \ \ \ \lambda_{Y_3}  = v^3w^3, \ \ \ \
 \lambda_B  = u^3v^2w^2, \ \ \ \ \lambda_C  = u^{3/2}v^3w^3
\ee
plus cyclic permutations of the triple $(u,v,w)$ for the $'$ and $''$ cases.
Like it is done in \cite{MMkrM}, ${\cal D}_{{\cal X}_3}$ denotes a sum of three dimensions, equal to
\be
{\cal D}_{{\cal X}_3} \equiv
{\cal D}_{X_3 }+{\cal D}_{X_3'}+{\cal D}_{X_3''} =
{\cal D}_{\Lambda^3(Adj)} - 1 - {\cal D}_{X_2} - {\cal D}_{Y_2(\alpha)} -
{\cal D}_{Y_2(\beta)} -{\cal D}_{Y_2(\gamma)}
\label{dimUX3}
\ee

\smallskip

\noindent
with ${\cal D}_{\Lambda^3(Adj)}(q) =
 \frac{{\cal D}_{Adj}(q)^3 - 3{\cal D}_{Adj}(q^2){\cal D}_{Adj}(q)+2{\cal D}_{Adj}(q^3)}{6}$.
 The three individual dimensions in this case are quite sophisticated even in the classical
 limit of $q=1$, but all the three have identical Casimirs and enter the formulas for
 knots as a sum, which has a long, but explicit expression through $u,v,w$.

\subsection{Obtaining arborescent knots}

Using the Racah matrix (\ref{uniRacah}) and the corresponding eigenvalues (\ref{eva}), one can immediately evaluate the
universal adjoint polynomials of the arborescent knots. Their expressions through this data can be found, e.g., in \cite{mmmrv}
or even extracted from the tables in \cite{Cau}. The final answers for the universal adjoint polynomials are quite long, hence, we do not list them here, but in order to illustrate how the procedure works we collected in \cite{knotebook2} the universal adjoint polynomials obtained in this way: for all arborescent knots with no more than 8 crossings (i.e. all knots with no more than 8 crossings except for single $8_{18}$ in the Rolfsen table, \cite{katlas}), and also for many arborescent knots with 9 and 10 crossings. Note that among these examples there are numerous knots that can be presented as well as 3,4,5,6-strand closed braids on one side, and knots that can be presented by double-fat graphs with up to 6 fingers, plenty of them being non-pretzel and even non-starfish ones on the other side:  the list is quite representative. However, to make the text readable, we discuss below the universal
adjoint polynomials for just two simple families of arborescent knots: for the twist and pretzel knots.

Before coming to these families, we make an important comment about the framing factor.
That is, despite there is no difference between the Racah matrices $U$, $S$ and $\bar S$ for the adjoint
representation, and between the sets of the corresponding eigenvalues (i.e. the universal adjoint knot polynomial does not depend on orientation, since the adjoint representation is contragredient), this is literally true only in group theory. For knots the situation is a little trickier. The point is that the group theory objects like universal ${\cal R}$-matrix correspond to the so called "vertical" framing in knot theory, which is different from the "topological" one.
In result, the topologically invariant knot polynomials, which we are looking for,
keep some memory about the difference between $S$ and $\bar S$, because their
transformations from vertical to topological framings are different.
This was the origin of additional factors in knot polynomials of the previous subsection (e.g. each ${\cal R}$-matrix in the previous subsection has to be multiplied by $t^4=(uvw)^4$ in order to obtain the correct formulas (\ref{3sk}), (\ref{3sl}) from (\ref{t0})-(\ref{t2}), and an additional factor $(uvw)^{4n}$ in (\ref{Padj2str}))
and this will produce some extra common powers of $t=uvw$ in our formulas below, they are the only memory of orientation dependence which survives in the universal sector of knot theory. We just insert these powers of $t$ wherever necessary without going into lengthy comments. In practice, say, for the pretzel knots, one picks up a representation from the tables
of \cite{gmmms} or \cite{mmmrv}, like $6_2=(3,\bar 2, 1)$ and inserts a factor $t^{-4}$
per each non-overlined item, while nothing for the overlined one, i.e.
a total of $t^{-4\cdot(3+1)} = t^{-16}$ for this realization of $6_2$
or $t^{-4\cdot(2-3+1-3)}=t^{12}$ for pretzel representation $(2,-3,1,-3)$ of $8_{21}$,
or $t^{-4\cdot(3+3-1-1)}=t^{-16}$
for pretzel representation $(\bar 4,3,3,-1,-1)$ of $10_{144}$, see formulas (\ref{pret}) and (\ref{genfinger}) below.

\subsection{Twist knots}

The twist knots are described by the knot diagrams like (here $k$ is equal to 3 and the knot is $7_2$):

\vspace{0.4cm}

\begin{figure}[h!]
\centering\leavevmode
\includegraphics[width=4.5cm]{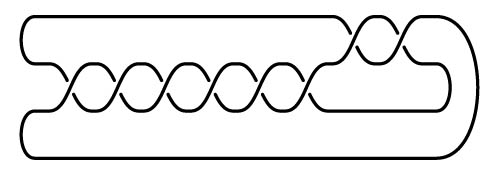}
\end{figure}

\noindent
The general formula for the universal adjoint polynomial for the twist knot is
\be
{\cal P}_{Adj}^{tw(k)} = {\cal D}_{Adj}\cdot\left({\cal S}\lambda^{-2k}{\cal S}\lambda^{-2}{\cal S}\right)_{11}={\cal D}_{Adj}\cdot \sum_{Q,Q'=1}^6 {\cal S}_{1Q}{\cal S}_{Q'1}{\cal S}_{QQ'}
\lambda_Q^{-2k}\lambda_{Q'}^{-2}
\label{Padjtw}
\ee
where $\lambda$ denotes the diagonal matrix with elements $\lambda_i$.

It can be represented in the general evolution formula \cite{evo} for the universal adjoint polynomial
for this family of knots
\be
{\cal P}^{tw(k)}_{Adj} = 1+ \sum_{j=2}^6 F_j(u,v,w)\cdot(\lambda_j^{-2k}-1)
\label{Ptwadj}
\ee
where
\be
\lambda_1=1, \ \
\lambda_2=tu^{-1}, \ \ \lambda_3 = tv^{-1}, \ \ \lambda_3=tw^{-1}, \ \
\lambda_5 = -t, \ \ \lambda_6=-t^2
\ee
with $t=uvw$ and the coefficients
\be
F_j^{twist} = {\cal D}_{Adj}\cdot{\cal S}_{1j}\cdot \sum_{k=1}^6{\cal S}_{jk}\lambda_k^{-2}{\cal S}_{k1}
\ee
 are
\be
F_2^{twist} = \frac{(t^2-1)(vt-1)(wt-1)(t^2-u^3)}{t^6(u-v)(u-w)(u+1)(t^2-u)}
\cdot\left(t^2-1+v-\frac{t^2}{v}+w-\frac{t^2}{w}\right),\nn \\
F_3^{twist} = \frac{(t^2-1)(ut-1)(wt-1)(t^2-v^3)}{t^6(v-u)(v-w)(v+1)(t^2-v)}
\cdot\left(t^2-1+u-\frac{t^2}{u}+w-\frac{t^2}{w}\right),\nn \\
F_4^{twist} = \frac{(t^2-1)(ut-1)(vt-1)(t^2-w^3)}{t^6(w-u)(w-u)(w+1)(t^2-w)}
\cdot\left(t^2-1+u-\frac{t^2}{u}+v-\frac{t^2}{v}\right),\nn \\
F_5^{twist} = -{(1+t)\over t^5}\left(t^2-1+u-\frac{t^2}{u}+v-\frac{t^2}{v}+w-\frac{t^2}{w}\right), \nn \\
F_6^{twist} = \frac{(1+t)(ut-1)(t+u)(vt-1)(t+v)(wt-1)(t+w)}{t^6(u+1)(v+1)(w+1)}
\ee
Due to orthogonality of ${\cal S}$ the sum $\sum_{j=1}^6 F_j^{twist} = 1$,
and this is already taken into account in (\ref{Ptwadj}),
where the sum goes from $j=2$.

This expression adds to the previously known evolution formulas for HOMFLY in all symmetric
representations from \cite{evo} and in representation $[21]$ from \cite{MMMtwist},
and the one for the fundamental Kauffman from \cite{II}.
Eq.(\ref{Ptwadj}) immediately reproduces those for the trefoil $tw(1)=3_1$
and figure eight knot $tw(-1)=4_1$, obtained by a tedious analysis in \cite{MMkrM}
and validates again somewhat risky arguments behind the "exotic terms" in the $4_1$ HOMFLY polynomials
inherited from \cite{Ano21}.

\subsection{Pretzel and other knots
\label{pretzel}}

One of the simplest subfamilies inside the arborescent knots is that of the pretzel links/knots,
their knot polynomials being analyzed in great detail in \cite{gmmms}.
If considered as made from the 2-strand braids, the pretzel link/knot looks like a $(g=k-1)$-loop
diagram naturally lying on the surface of genus $g$:

\begin{figure}[h!]
\centering\leavevmode
\includegraphics[width=7cm]{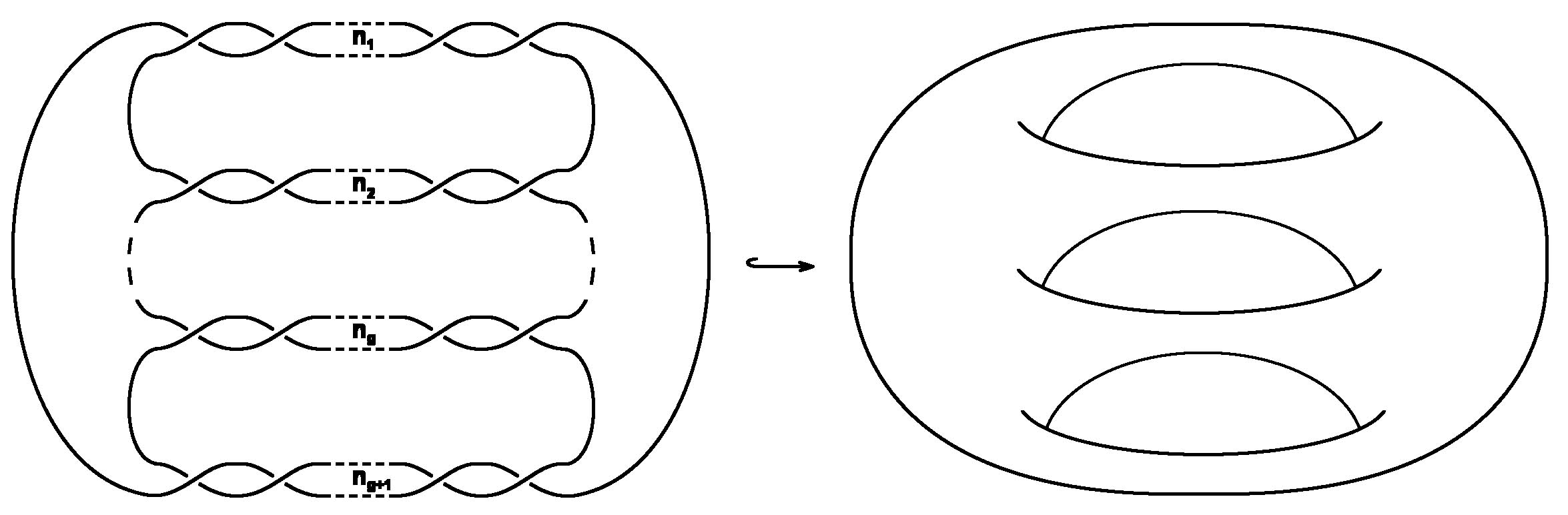}
\end{figure}

Knot polynomials for pretzel knots are made from "pretzel fingers"
by the simple rule \cite{gmmms}:
\be
{\cal P}^{pretzel(p_1,\ldots,p_n)}_{Adj} =
{\cal D}_{Adj}\cdot \sum_{j=1}^6 {\cal S}_{1j}^{2-n} \prod_{m=1}^n F_j^{pretzel}(p_m)
\ee
where fingers are described by their own evolution formula:
\be\label{pret}
F_j^{pretzel}(p) \equiv (uvw)^{4N(p)}\cdot \sum_{k=1}^6 {\cal S}_{jk}\lambda_k^{-p} {\cal S}_{k1} =
(uvw)^{4N(p)}\cdot\sum_{k=1}^6 \lambda_k^{-p} F_{j,k}^{pretzel}
\ee
where $N(p)$ is equal to $p$ for "parallel fingers" (i.e. those made of parallel 2-strand braid) and 0 for antiparallel.

For $n=1$ we get just an unknot, for $n=2$ the 2-strand knots/links $[2,p_1+p_2]$,
less trivial examples begin from $n=3$, see tables of pretzel knots in
the second paper of \cite{gmmms} and in \cite{mmmrv}.

As usual, the situation in the adjoint (universal) sector is simpler
than in \cite{gmmms}, because one does not need to pay attention to
orientation and differences between $S,S^\dagger$ and $\bar S$,
and also we get the answers for all groups at once.
Note that, since every of the six representations appears in $Adj^{\otimes 2}$
just once, each $F_j$ is {\it not} a matrix, there are no extra indices
$a,b=1,\ldots, dim(W_j)=1$, see \cite{mmmrv}; therefore adjoint
polynomials do not distinguish pretzel mutants (not only antiparallel!),
just as the $[21]$-colored HOMFLY did for $SU(3)$
(however, many of them were separated by $H_{[21]}$ for $N>3$).

The same formula with more complicated fingers, each of them being a 4-strand braid
\be
F_j^{(p_1,q_1,\ldots)} = t^{4N(p_1,q_1,...)}\cdot ({\cal S}\mu^{p_1}{\cal S}\mu^{q_1}{\cal S}\mu^{p_2}\ldots {\cal S})_{j1}
\label{genfinger}
\ee
describes {\it starfish} knots/links \cite{mmmrv} that look like (3-finger example)

\begin{figure}[h!]
\centering\leavevmode
\includegraphics[width=3.3 cm]{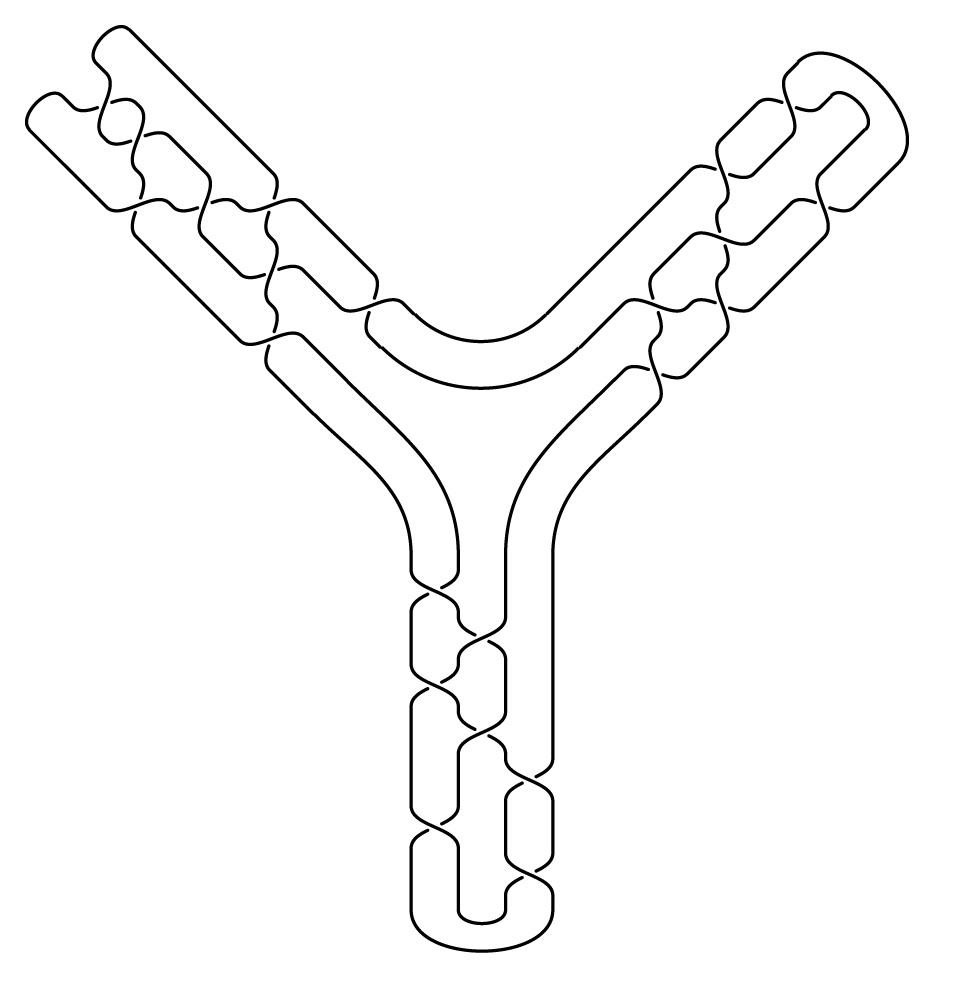}
\label{example}
\end{figure}

The 2-bridge knots correspond to the case of two arbitrary fingers, which is in fact the same as one arbitrary and one trivial finger.

Generic arborescent knots/links are starfish configurations, connected by the
double-fat "propagators"
\be
\Pi_{jk}^{(p_1,q_1,\ldots)} = ({\cal S}\mu^{p_1}{\cal S}\mu^{q_1}{\cal S}\mu^{p_2}\ldots {\cal S})_{jk}
\label{genprop}
\ee
to form "starfish trees" \cite{mmmrv}: loops are not allowed.
Finger (\ref{genfinger}) is the propagator (\ref{genprop}) with one "vacuum" index $k=1=\emptyset$). The typical configuration looks like (one propagator and four fingers)

\begin{figure}[h!]
\centering\leavevmode
\includegraphics[width=4.5cm]{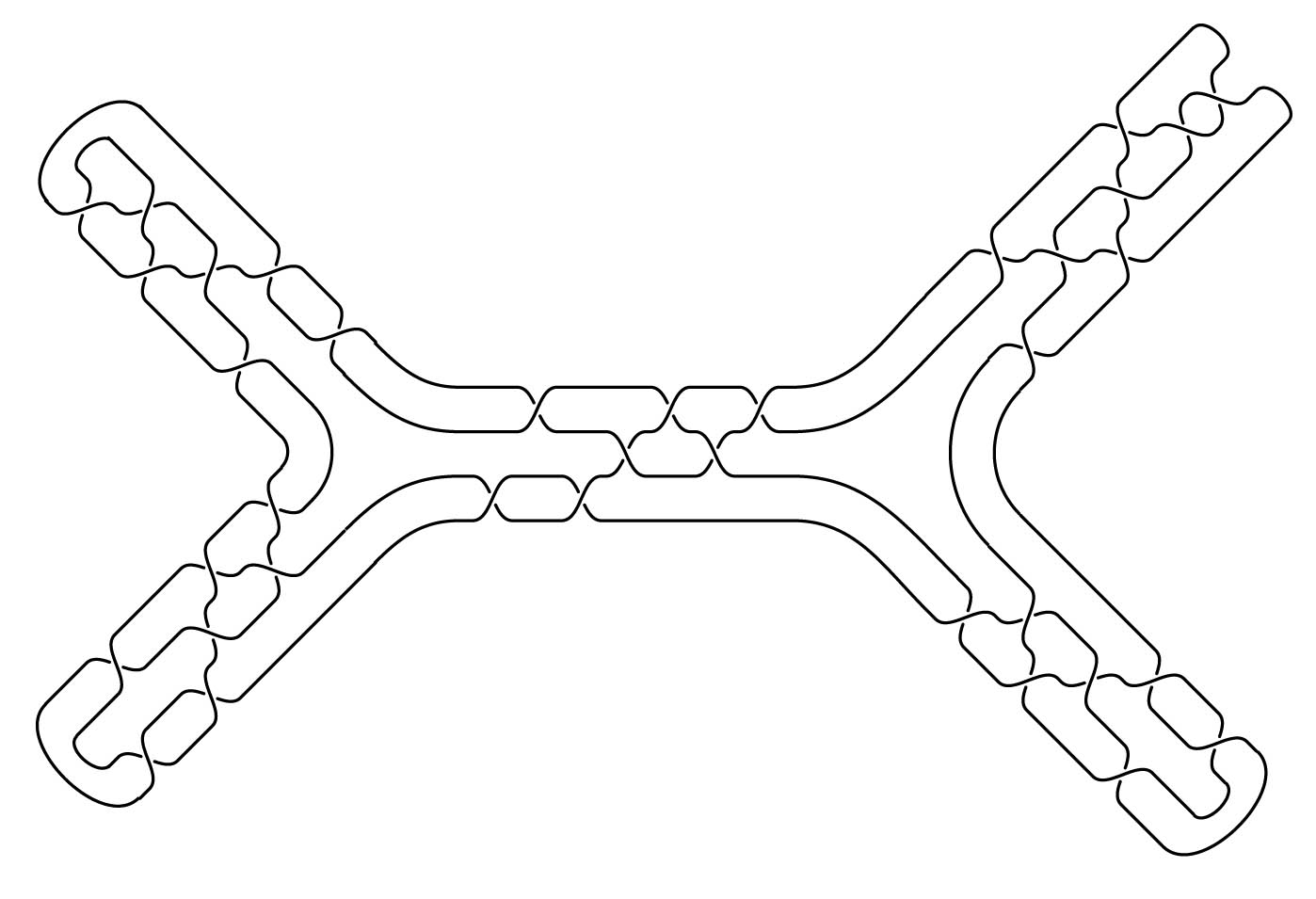}
\label{braidg}
\end{figure}

\subsection{Mutants}

Starting from the pretzel family there are plenty of mutant knots
(though the simplest 11-intersection mutants are non-pretzel but still
arborescent, see \cite{nrv} and \cite{mmmrv}  for their separation by
$[21]$-colored HOMFLY).
Though the adjoint representation is not symmetric, mutants could not be
separated by our universal adjoint polynomials for arborescent knots (not only for the pretzel ones), for the reason that
we already mentioned in s.\ref{pretzel}: each representation appears in $Adj^{\otimes 2}$ only once and
fingers are {\it not} matrices. This is definitely in agreement with the claim of \cite{Morton,mmmrv,nrv} that the arborescent mutants could be distinguished in representation $[21]$ only for the groups $SU(N)$ with $N>3$. However, even this is not enough: antiparallel pretzel mutants are not distinguished by $[21]$-colored HOMFLY at all, at least the representation $[42]$ is needed \cite{Morton2}.
At the same time, descendants of the adjoint sometimes already have non-trivial multiplicities in their square:
for instance, the tensor square of the representation $X_2$ that emerged in the tensor square of the adjoint contains an irrep with multiplicity 2. Hence, one may hope to separate different mutants with universal knot polynomials evaluated in these
higher representations.

\section{Comments and discussion}

\subsection{Three-strand calculations beyond arborescent knots}

In fact, one can now move further to go beyond the arborescent family. For instance, one can consider the generic (not obligatory torus) knot/link given by a 3-strand closed braid, which is described by the second formula (\ref{3st}), only the expression under the trace is more complicated: $\prod_i {\cal R}^{a_i}{\cal S}{\cal R}^{b_i}{\cal S}$,
where the sets $\{a_i\}$ and $\{b_i\}$ can contain not only unities. Since the matrices ${\cal S}$ are now explicitly known from our extension of the eigenvalue conjecture to $N=6$, this is a straightforward exercise to evaluate any $3$-strand closed braid. However, in the present paper we choose another direction and describe the complementary story  of how the knowledge of ${\cal S}$ provides answers for the arborescent knots. Then, the mixture of the 3-strand and arborescent calculi provides the technique to handle the knot/link diagrams given by "the fingered 3-strand closed braids" of \cite{MM3f1,MM3f2}, the biggest family by now allowing systematic evaluation of colored knot polynomials. These diagrams look like

\unitlength 0.30mm 
\linethickness{0.5pt}
\begin{picture}(300,100)(-70,-50)
\put(0,30){\vector(1,0){35}}
\put(0,0){\vector(1,0){35}}
\put(0,-30){\vector(1,0){35}}
\put(30,-30){\vector(1,0){125}}
\qbezier(35,30)(40,30)(40,25)
\qbezier(35,0)(40,0)(40,5)
\put(35,25){\line(1,0){20}}
\put(35,5){\line(0,1){20}}
\put(40,12){\mbox{$F_1$}}
\put(35,5){\line(1,0){20}}
\put(55,5){\line(0,1){20}}
\qbezier(50,25)(50,30)(55,30)
\qbezier(50,5)(50,0)(55,0)
\put(55,30){\vector(1,0){20}}
\put(55,0){\vector(1,0){20}}
\qbezier(75,30)(80,30)(80,25)
\qbezier(75,0)(80,0)(80,5)
\put(75,25){\line(1,0){20}}
\put(75,5){\line(0,1){20}}
\put(80,12){\mbox{$F_2$}}
\put(75,5){\line(1,0){20}}
\put(95,5){\line(0,1){20}}
\qbezier(90,25)(90,30)(95,30)
\qbezier(90,5)(90,0)(95,0)
\put(95,30){\vector(1,0){20}}
\put(95,0){\vector(1,0){20}}
\qbezier(115,30)(120,30)(120,25)
\qbezier(115,0)(120,0)(120,5)
\put(115,25){\line(1,0){20}}
\put(115,5){\line(0,1){20}}
\put(120,12){\mbox{$F_3$}}
\put(115,5){\line(1,0){20}}
\put(135,5){\line(0,1){20}}
\qbezier(130,25)(130,30)(135,30)
\qbezier(130,5)(130,0)(135,0)
\put(135,30){\vector(1,0){20}}
\put(135,0){\vector(1,0){20}}
\qbezier(155,-30)(160,-30)(160,-25)
\qbezier(155,0)(160,0)(160,-5)
\put(155,-25){\line(1,0){20}}
\put(155,-5){\line(0,-1){20}}
\put(160,-17){\mbox{$F_4$}}
\put(155,-5){\line(1,0){20}}
\put(175,-5){\line(0,-1){20}}
\qbezier(170,-25)(170,-30)(175,-30)
\qbezier(170,-5)(170,0)(175,0)
\put(155,30){\line(1,0){40}}
\put(175,-30){\vector(1,0){60}}
\put(175,0){\vector(1,0){20}}
\qbezier(195,30)(200,30)(200,25)
\qbezier(195,0)(200,0)(200,5)
\put(195,25){\line(1,0){20}}
\put(195,5){\line(0,1){20}}
\put(200,12){\mbox{$F_5$}}
\put(195,5){\line(1,0){20}}
\put(215,5){\line(0,1){20}}
\qbezier(210,25)(210,30)(215,30)
\qbezier(210,5)(210,0)(215,0)
\put(215,30){\vector(1,0){70}}
\put(215,0){\vector(1,0){20}}
\qbezier(235,-30)(240,-30)(240,-25)
\qbezier(235,0)(240,0)(240,-5)
\put(235,-25){\line(1,0){20}}
\put(235,-5){\line(0,-1){20}}
\put(240,-17){\mbox{$F_6$}}
\put(235,-5){\line(1,0){20}}
\put(255,-5){\line(0,-1){20}}
\qbezier(250,-25)(250,-30)(255,-30)
\qbezier(250,-5)(250,0)(255,0)
\put(255,-30){\vector(1,0){30}}
\put(255,0){\vector(1,0){30}}
%
\end{picture}

\noindent
where $F_i$ are fingers. This diagram is described by the formula
\be
\sum_{Q\in Adj^{\otimes 3}} \frac{{\cal D}_Q}{{\cal D}_{Adj}}\cdot
\Tr_Q (F_1F_2F_3{\cal S}F_4{\cal S} F_5{\cal S} F_6{\cal S})_Q
\label{fingered3strand}
\ee
which differs from generic 3-strand braid expressions by substitution of powers of ${\cal R}$-matrices
by arbitrary double-fat fingers (\ref{genfinger}). In fact, one is allowed to insert arbitrary fingers at any place of the diagram.
Further generalization connects such 3-strand loops by propagators (\ref{genprop})
to form fingered-loops trees (no loops!).
To evaluate these quantities, one needs to know mixing matrices and fingers in all irreps in
$Q\in Adj^3$.
The fingers are actually made from the same mixing matrices (interpreted as Racah matrices),
thus the issue is only these matrices.
In the 3-strand case, the maximal size of $W_Q$ is $6$, and it appears in just two cases:
for $Q=X_1=Adj$ and for $Q=X_2$. The former case is exhaustively discussed in the present
paper, thus the item with $Q=X_1$ in (\ref{fingered3strand}) can be evaluated.
All other terms with $Q\neq X_1,X_2$ involve mixing matrices of smaller sizes,
available from \cite{IMMMev}.
Thus, what remains is the $6\times 6$ matrix for $X_2$,
which should be extractable  from the general expression in s.\ref{mima}.

Another delicate point is to handle three representations $X_3\ X_3',\ X_3''\in Adj^{\otimes 3}$. They have uncomprehensible dimensions, even classical ones \cite[Theorem 3.8]{V2}, however, the knot polynomials seem to depend only on the sum of them, which is simple. This property, however, requires a careful check.

Thus it remains to list mixing matrices, fingers and traces for all $Q \in Adj^{\otimes 3}$,
this is a straightforward, but tedious work, and the results require quite some space
to be exposed. They will be reported in a sequel paper \cite{II}, and we include there also further details about the 3-strand calculus.

\subsection{Properties of the universal adjoint polynomials}

In all new examples the properties, observed in \cite{MMkrM} continue to hold:

\paragraph{i)} special polynomial property at $u=v=1$:
\be
{\cal P}_{Adj}(u=v=1,w=A) = \Big(H_{_\Box}(q=1,A)\Big)^2 =
\Big(K_{_\Box}^2(q=1,A)\Big)^2
\label{spepoP}
\ee
\paragraph{ii)} differential expansion property:
\be
{\cal P}_{Adj}(u,v,w ) -1 \ \ \vdots \ \ (uvw-1)(uvw+1)
\ee
in particular, at $t=uvw=\pm 1$ (i.e. at $w=\pm\frac{1}{uv}$:
\be
{\cal P}_{Adj}(u,v,w)\Big|_{uvw=\pm 1} =1
\ee
\paragraph{iii)} at $w=\pm 1$ additionally
\be
{\cal P}_{Adj}(w=1,u,v)-1 \ \ \vdots \ \ (uv-1)^2(uv+1)
\ee
and
\be
{\cal P}_{Adj}(w=-1,u,v)-1 \ \ \vdots \ \ (uv-1)^2(uv+1)
\ee
note that the extra factor $(uv-1)$ is the same (with minus) in both cases.

\section{Conclusion \label{conc}}

In \cite{MMkrM} the notion of universal knot polynomials was introduced. In this paper we described the universal adjoint polynomials,  for a huge variety of arborescent knots providing a solid ground for the claims of \cite{MMkrM}. In fact, in \cite{MMkrM} the main  examples were torus knots, and these could be described by the Rosso-Jones formula. Since it is made out of quantum dimensions and Casimirs, its universalization is almost obvious:   there is nothing "knotty" in it.
The value of \cite{MMkrM} was that the emphasize was put on the
arguments that can have more general applications, beyond the torus knots.
As a single illustration, an example of $4_1$ was considered, and the corresponding universal adjoint knot polynomial was guessed by {\it interpolation} of the uniform adjoint HOMFLY and Kauffman polynomials.

In this paper we provided another, missed ingredient to story: using the eigenvalue conjecture of \cite{IMMMev} we defined and evaluated {\it the universal Racah matrix} ($6j$-symbol) for the four adjoint representations. This allows one to {\it evaluate} the universal adjoint knot polynomials of any arborescent knots. In particular, in this was we confirmed that the polynomials obtained in \cite{MMkrM} are correct.

This scheme can be directly extended to non-arborescent knots: this just requires more mixing matrices in the universal form. We
illustrate this fact with the important example of fingered 3-strand knots/links in the next paper of this series, \cite{II}.

\section*{Acknowledgements}

We are indebted to R. Mkrtchyan who introduced us to the wonderful world of universality and generously shared with us his deep knowledge of this field.

This work was performed at the Institute for Information Transmission Problems with the financial
support of the Russian Science Foundation (Grant No.14-50-00150).

\end{document}